\documentclass[]{spie}  

\usepackage{amsmath,amsfonts,amssymb}
\usepackage{graphicx}
\usepackage[colorlinks=true, allcolors=blue]{hyperref}
\usepackage{subcaption}
\usepackage{tikz}
\usepackage{booktabs}
\usepackage{siunitx}
\title{Theoretical Analysis of Fiber Arrangements for Improved Coupling Efficiency in the VBT High-Resolution Echelle Spectrograph}

\author[a]{Nitish Singh}
\author[a]{S. Sriram}
\author[a]{Bharat Kumar Yerra}
\author[a]{Prasanna G. Deshmukh}
\affil[a]{Indian Institute of Astrophysics, II Block, Koramangala, Bengaluru 560 034, INDIA}

\authorinfo{Further author information: Residence: Bhaskara, Indian Institute of Astrophysics Guest House, Koramangala II Block, Bengaluru 560 034, India\\
Nitish Singh: E-mail: nitish.singh@iiap.res.in, nitiphy@gmail.com\\
Telephone: +91 8077759357}

\begin{document} 
\maketitle

\begin{abstract}
In fiber-based spectroscopy within telescopes, a prevailing limitation has been the necessity to align the fiber diameter with the telescope’s seeing conditions, often characterized by the Full Width at Half Maximum of the point spread function. This alignment constraint captures around 50 \% of the incoming flux from any point source. Furthermore, the challenge is compounded when high-resolution spectroscopy is in play, as it often demands a minute slit width, further exacerbating flux loss. The essence of this paper lies in a comprehensive exploration, accomplished through theoretical simulations, of strategies aimed at enhancing the coupling efficiency of high-resolution spectrographs. The primary objective is to bolster the flux capture without compromising the critical aspect of spectral resolution. This research endeavors to unlock the potential for more effective utilization of high-resolution spectrographs to study celestial objects.

\end{abstract}


\section{INTRODUCTION}

\begin{figure}[htbp]

    \centering
    \includegraphics[width=0.9\linewidth]{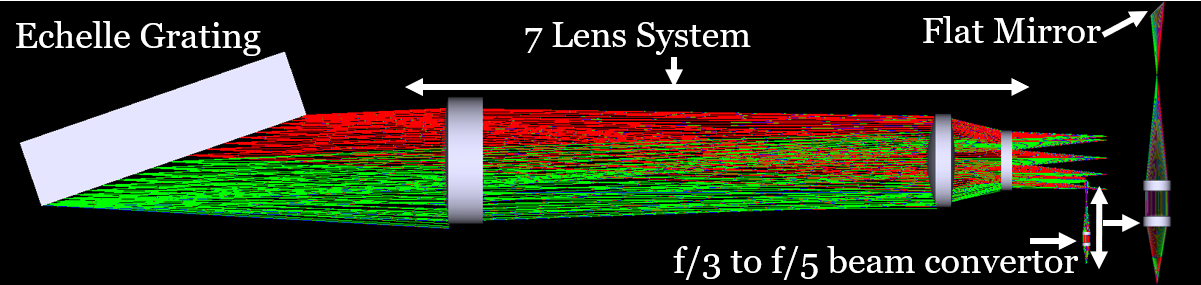}
    \caption{Optical Design of HRES}
    \label{fig:echelle_layout}

\end{figure}

The 2.34m Vainu Bappu Telescope (VBT) is a reflecting telescope located at Vainu Bappu Observatory in India. The High-Resolution Echelle spectrograph (HRES) is presently operating at the prime focus of the VBT. It is designed for a f/5 beam and capable of achieving spectral resolutions (R) $\approx$ 27,000 - 100,000, (\citenum{2005JApA...26..331R}). The prime focus of the telescope has the F-number of f/3.25, which results in an image scale of 27 arcsec/mm. The HRES allows light from the prime focus to be fed through a f/3 optical fiber, typically with a fixed diameter. The seeing conditions at the VBT site vary between 2 and 3.5 arcsec, with an average seeing of 2.5 arcsec (\citenum{2009MNRAS.395..593P}). This is particularly notable because the diameter of the fiber is typically fixed, leading to an imperfect match with instantaneous seeing conditions (\citenum{1986AJ.....92..219H}). Consequently, the core diameter is often around 2.7 arcsec or 100 µm. This alignment is crucial as it directly affects the amount of incoming light captured from celestial sources, with the current fiber typically capturing only around 55 \% of the available flux. The light from the fiber output, which forms an f/3 beam, is directed through f/3 to f/5 converging lenses to meet the beam requirements of the HRES. At 80 mm from the second lens, a fold mirror redirects the light toward a set of seven lenses configured with an f/5 system. These lenses, with an effective focal length of 755 mm and a diameter of 151 mm, play a crucial role in further focusing and shaping the light. Subsequently, a grating introduces dispersion, directing the light along the incoming beam's trajectory at an angle. Finally, a detector, positioned 100 mm away from the seven-lens system and 30 mm from the initial fold mirror, captures the spectra of dispersed light (\citenum{2005JApA...26..331R}). The optical layout of the HRES is illustrated in Fig.~\ref{fig:echelle_layout}.  A 4k $\times$ 4k CCD is utilized in the HRES, where each pixel corresponds to 12 $\mu m$.

The HRES provides a range of spectral resolutions by adjusting the slit width, such as R $\approx$ 27,000 with the full slit open, and even higher resolutions like R $\approx$ 72,000 and R $\approx$ 100,000 with narrower slit widths of $60\mu m$ and $30\mu m$, respectively (\citenum{2005JApA...26..331R}). Maintaining high throughput for higher spectral resolution due to substantial light loss through the slit poses a challenge. In this paper, we discuss various core-shaped fiber arrangements (circular, rectangular, and hexagonal-shaped fibers) to evaluate their throughput using simulations (\citenum{2021JPhCS1717a2048N}). Our detailed theoretical analysis provides valuable insights into optimizing light capture for high-resolution fiber-fed spectrograph setups.

\section{Analysis of Throughput Estimation and Flux Calculation of Current HRES Setup}

\label{sec:current_setup}

\begin{figure}[htbp]
  \centering
  \begin{subfigure}[h]{0.45\linewidth}
    \centering
    \includegraphics[width=\linewidth]{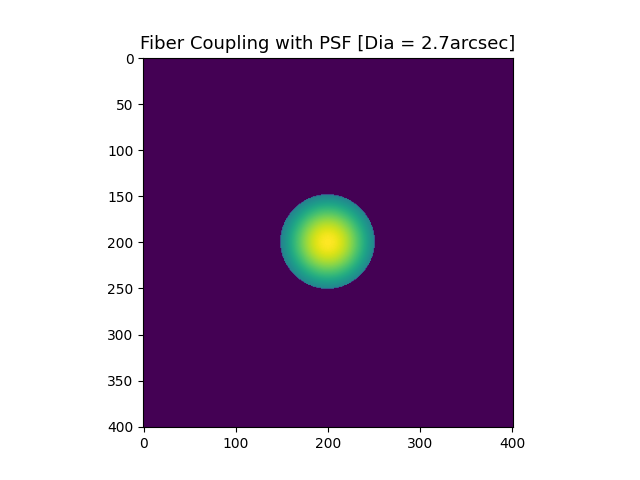}
    \caption{2.7 arcsec core diameter fiber coupling}
    \label{fig:100um_fiber}
  \end{subfigure}
  \hspace{5mm}
  \begin{subfigure}[h]{0.45\linewidth}
    \centering
    \includegraphics[width=\linewidth]{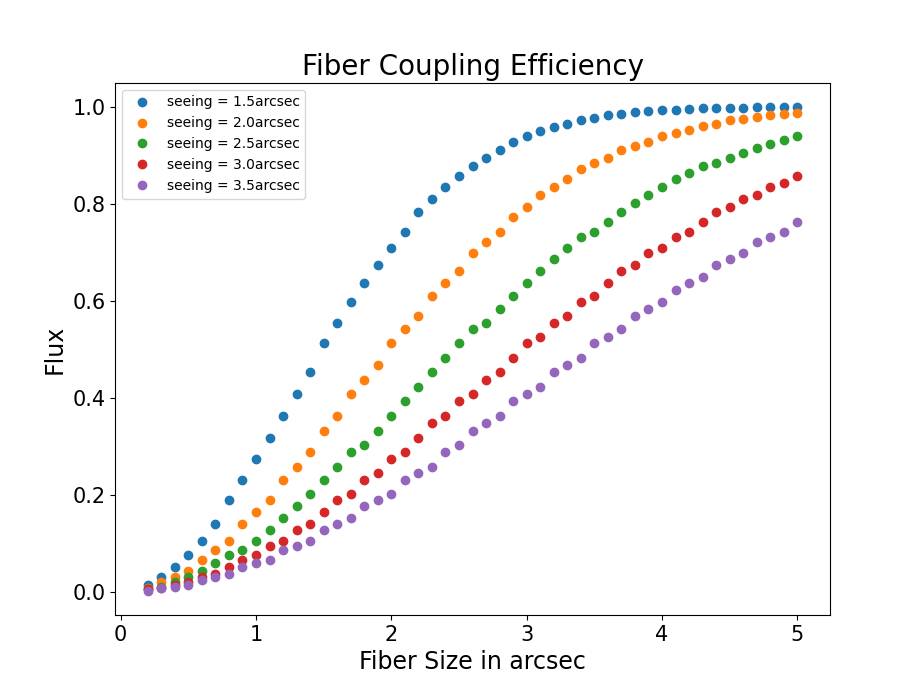}
    \caption{Fiber coupling for different seeing or different diameter fiber}
    \label{fig:100um_fibercoup}
  \end{subfigure}
  \caption{Illustrations of fiber coupling and different fiber configurations}
  \label{fig:fiber_coupl}
  \vspace{-3mm}
\end{figure}

Currently, HRES operates at the VBT prime focus, fed by a circular core optical fiber with a 100µm core diameter. The Point Spread Function (PSF) induced by atmospheric conditions initially manifests as a speckle pattern during short exposures, evolving into a 2D Gaussian distribution with longer exposures (\citenum{1969A&A.....3..455M}). The Full Width at Half Maximum (FWHM) of the PSF aligns with the telescope’s seeing (\citenum{2023FrASS..1058213L}). In the simulations, We generate a 2D Gaussian PSF with different seeing conditions, which is subsequently coupled with a 2.7 arcsec (or $100 \mu m$) fiber to calculate the total throughput (see Fig.~\ref{fig:100um_fiber}). The seeing at VBT varies between 2.5 and 3.5 arcsec, We calculated the fiber coupling efficiency for different seeing conditions, as shown in Fig.~\ref{fig:100um_fibercoup}. For an average seeing of 2.5 arcsec with a 2.7 arcsec core diameter fiber, the coupling efficiency is 55\% which indicates around 45 \% of light loss due to aperture loss in prime focus.

\vspace{-3mm}

Other than fiber coupling, the optics used in the HRES system introduced reflection and transmission flux losses  which are given by:

Throughput Analysis of HRES Optical System = Primary Mirror Reflectivity $\times$ f/ratio converter Transmissivity $\times$ Collimator Lens Transmissivity $\times$ Camera Lens Transmissivity $\times$ Grating Reflectivity
    \begin{equation}
       \text{Throughput by HRES Optics} = 0.80 \times (0.98)^2 \times (0.98)^7 \times (0.98)^7 \times 0.98 = 0.58 \text{ or } 58 \%
       \label{eq:optics_thro}
    \end{equation}

Throughput by HRES Optics and Fiber Coupling = Throughput by HRES Optics $\times$ Fiber Coupling on Prime focus $\times$ Fiber Transmissivity
    
    \begin{equation}
       \text{Throughput by HRES Optics and Fiber Coupling} = 0.58 \times 0.55 \times 0.98 = 0.31 \text{ or } 31 \%
       \label{eq:opt_fib_thro}
    \end{equation}
    
In Equation \ref{eq:optics_thro}, We calculated the flux loss caused by the optics used in HRES. In Equation \ref{eq:opt_fib_thro} describes the total throughput of the HRES optics, including fiber coupling at the prime focus. We have taken maximum transmittance and reflectance into account, considering 98 \% for all lenses and the grating. Additionally, a reflectivity of 80 \% has been taken into account for the primary mirror of VBT. So in Equation \ref{eq:opt_fib_thro}, the throughput calculation considers fiber coupling, primary mirror reflectivity, and the optical elements used in HRES. However, slit loss is not included in this calculation, as discussed in Table \ref{tab:table1}. Without accounting for slit loss, the total throughput is only 31 \%.

\vspace{-4mm}
\begin{figure}[htbp]
  \centering
  \begin{subfigure}[h]{0.47\linewidth}
    \centering
    \includegraphics[width=\linewidth]{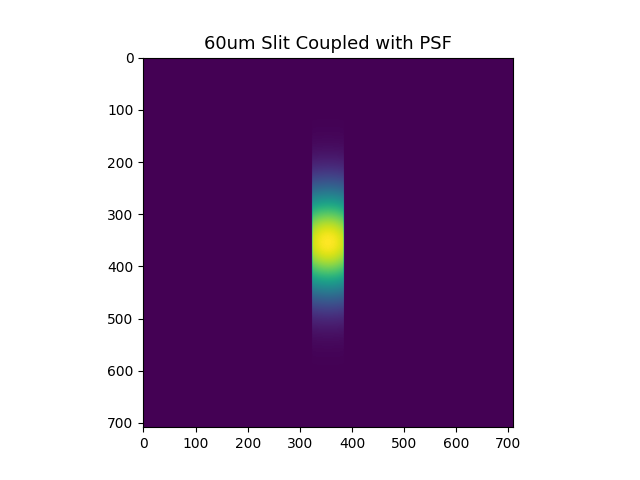}
    \caption{60 $\mu$m slit coupling after converting fiber output PSF f/3 beam to f/5 beam}
    \label{fig:100u_slit_coup}
  \end{subfigure}
  \hspace{5mm}
  \begin{subfigure}[h]{0.47\linewidth}
    \centering
    \includegraphics[width=\linewidth]{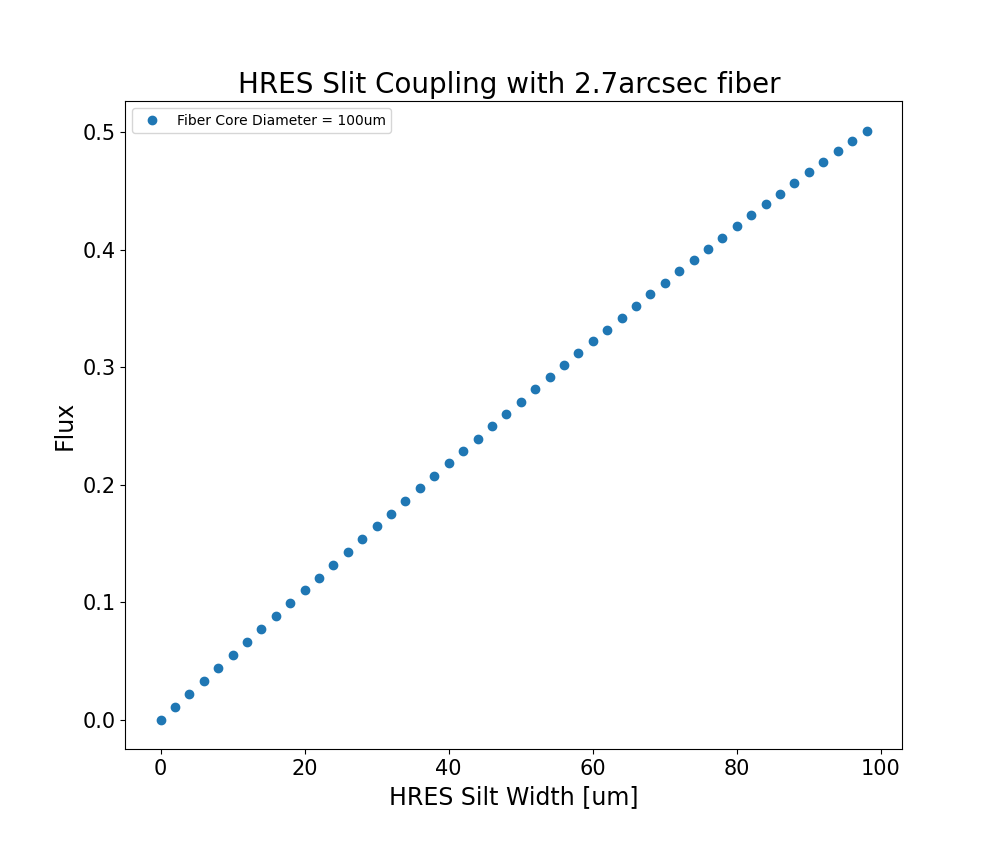}
    \caption{Different slit widths coupling with 100 $\mu m$ Fiber}
    \label{fig:100u_slit_coup_graph}
  \end{subfigure}
  \caption{Illustrations of slit coupling and different slit widths}
  \label{fig:slit_coupl}
  \vspace{-3mm}
\end{figure}

 Currently, the fiber output features an f/3 beam. Fiber output power distribution also follows a Gaussian form, where the FWHM corresponds to the core diameter of the fiber (\citenum{2002PASP..114..866R}). When coupling this power distribution function with different slit widths, further flux loss occurs by slit jaws see Fig.~\ref{fig:slit_coupl}. However, other optics utilized within the spectrograph contribute to additional flux loss. Table \ref{tab:table1} illustrates the flux loss calculations for the three operational modes of the HRES.

 \begin{table}[h]
    \centering
    \caption{\textbf{Total Throughput on the Detector}}
    \begin{tabular}{|p{2.3cm}|p{2cm}|p{2cm}|p{3.5cm}|p{3cm}|}
        \hline
        \textbf{Resolution (R)} & \textbf{Slit Width $(\mu m)$}  & \textbf{Slit Coupling}  & \textbf{Throughput by HRES Optics and Fiber Coupling} & \textbf{Total Throughput to detector}  \\
         \hline
        27,000 & Full slit open  & 0.99  & 0.31& 0.30    \\
        \hline
        72,000 & 60  & 0.33  & 0.31   & 0.10   \\
        \hline

        100,000 & 30  & 0.17  & 0.31   & 0.053    \\
        \hline
    \end{tabular}
    \label{tab:table1}
\end{table}

\begin{figure}[htbp]
  \centering
  \begin{subfigure}[b]{0.45\linewidth}
    \includegraphics[width=\linewidth]{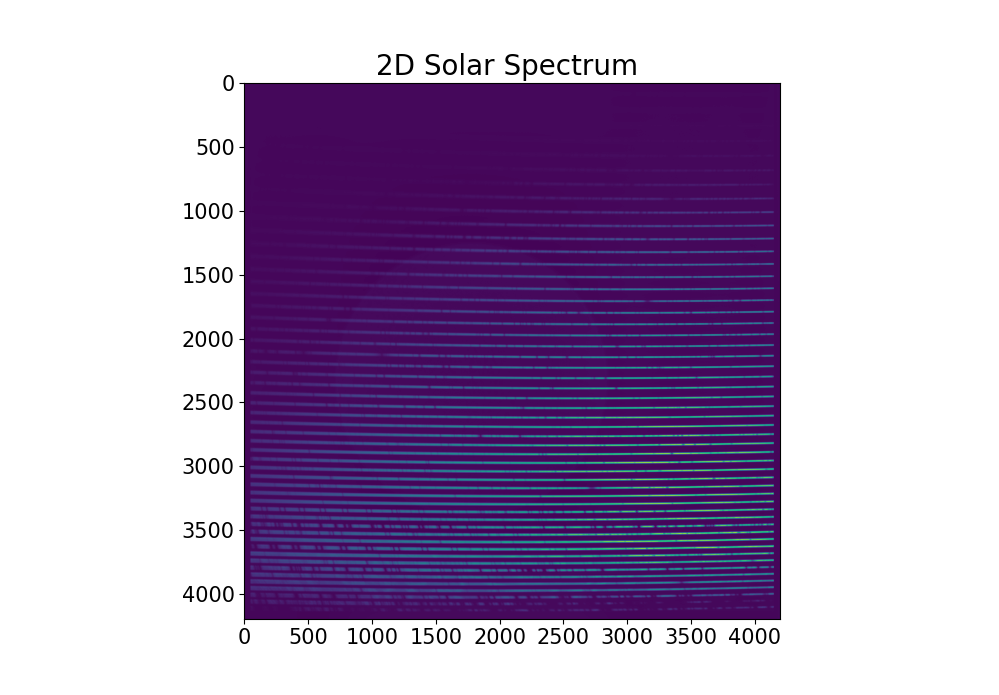}
    \caption{2D Solar Spectrum From HRES}
    \label{fig:2d_spectra1}
  \end{subfigure}
  \begin{subfigure}[b]{0.54\linewidth}
    \includegraphics[width=\linewidth]{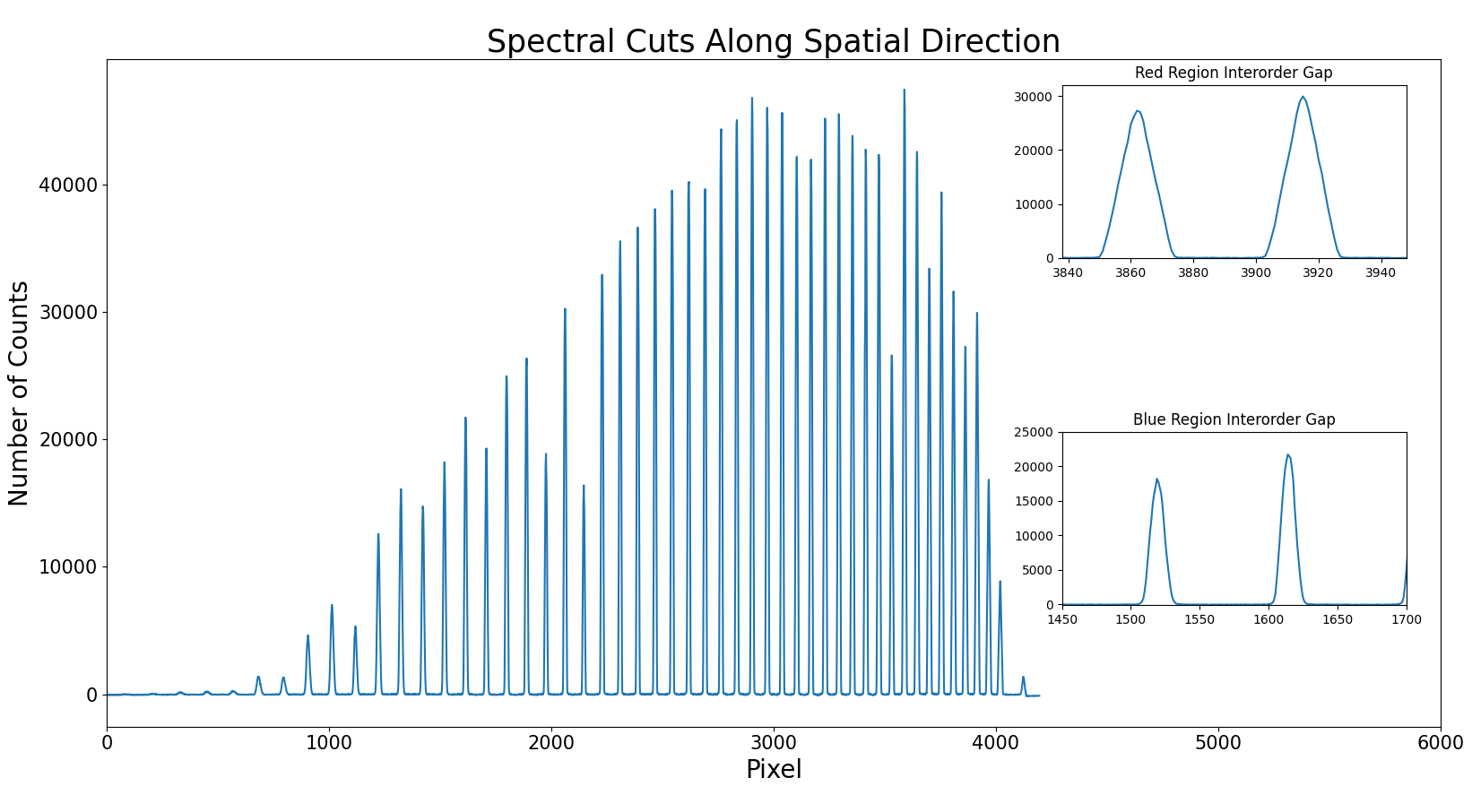}
    \caption{Spectral Cuts Along Spatial Direction}
    \label{fig:2d_spectra2}
  \end{subfigure}
  \caption{2D Solar Spectra Image Analysis}
  \label{fig:2d_spectra}
\end{figure}

The flux entering the spectrograph is notably low, with only 10 \% and 5 \% for R $\approx$ 72,000 and 100,000, respectively, as outlined in Table \ref{tab:table1}. Such substantial light loss underscores the need for alternative approaches. When considering alternative methods, it's crucial to examine the inter-order separation between consecutive orders to ensure sufficient separation. This will allow to avoid the overlap of spectral orders. To do this, we analyzed solar spectrum data captured by HRES across various resolutions and observed the inter-order separation between two orders, as illustrated in Fig.~\ref{fig:2d_spectra}. Notably, the inter-order separation is more pronounced at blue channel wavelengths compared to red channel wavelengths. For example, at red channel wavelengths, the inter-order separation is around 24 pixels, while at blue channel wavelengths, it is around 60 pixels. Each order width in the spatial direction measures approximately 32 pixels on average. This separation is influenced by the illuminated slit height due to the fiber flux. Consequently, to ensure adequate separation between adjacent inter-orders and prevent overlap, we aim to employ a fiber configuration with a height, after magnification, of 32 pixels or less. This adjustment will help maintain sufficient separation between adjacent inter-orders across the spectrum.

\vspace{-3mm}
\vspace{-3mm}
Hence, it's crucial to configure the fiber into a slit shape, ensuring that the height of the slit image on the CCD remains below 32 pixels for red channel wavelengths and 60 pixels for blue channel wavelengths. Overall, to maintain sufficient inter-order separation, the slit image height should be less than 32 pixels. Hence, to mitigate these constraints, alternative strategies are being considered to enhance light collection efficiency and mitigate losses, especially in scenarios demanding higher resolutions. This becomes particularly relevant as reducing the slit width to achieve higher resolution becomes less viable due to the excessive loss of light.

\section{Alternative Methods Explored for Enhancing Coupling of HRES}

In our pursuit of improved coupling efficiency, we are considering employing a bundle of multiple core-shaped fibers in prime mode to ensure that our inter-order separation and spectral resolution remains uncompromised see Fig.~\ref{fig:fiber arrangment}. In this scenario, the output from the fibers is in the form of an f/3 beam, which needs to be transformed into an f/5 beam to align with the requirements of the HRES, so the magnification factor is 1.6. Our theoretical target for achieving good inter-order separation is that the height of each fiber should ideally remain below 32 pixels (as calculated in section \ref{sec:current_setup} (see Fig.~ \ref{fig:2d_spectra})), which is equivalent to 384 $\mu$m since each pixel on the HRES CCD is $12\mu$m. So height of a single fiber is given by:

\vspace{-3mm}
\begin{equation}
\text{Height of a single fiber}  = \frac{\text{Width of order}}{\text{Number of Fiber} \times \text{Magnification factor}}
\label{eq:height_fiber}
\end{equation}
\subsection{Circular Fiber}

We plan to use 7 circular fibers arranged in a focal plane, as shown in Fig.~\ref{fig:sub1}. The output from these fibers will form a slit for the spectrograph. Considering the magnification required for transforming the fiber output, we calculate the diameter of each fiber using Equation.~\ref{eq:height_fiber}:

\vspace{-3mm}
\begin{equation}
    \text{Fiber diameter (core + cladding)} = \frac{384}{7 \times 1.6} = 34\mu m
    \label{eq:7circuler}
\end{equation}
From Equation \ref{eq:7circuler}, we find that the maximum fiber diameter we can use in our simulation is $34\mu m$. Therefore, we will use a $32\mu m$ core diameter with a $2\mu m$ cladding diameter fiber.

\subsection{Rectangular Fiber}

We are using three Rectangular fibers arranged in a focal plane, as illustrated in Fig.~\ref{fig:sub2} in our simulation, and we calculate the height of each fiber using Equation.~\ref{eq:height_fiber}: 

\vspace{-3mm}    
\begin{equation}
    \text{Fiber Height (core + cladding)} = \frac{384}{3 \times 1.6} = 80\mu m
    \label{eq:7rect}
\end{equation}

From Equation \ref{eq:7rect}, we find that the maximum fiber diameter we can use in our simulation is $80 \mu m$. Therefore, we will use a Each fiber has a height of $78\mu m$, with a cladding of $2\mu m$. Our aim is to achieve a fiber output diameter of approximately $60\mu m$ after magnification, facilitating direct feeding into the spectrograph to achieve $R = 72,000$. Therefore, we have chosen a core width of $37\mu m$ for the fiber (see Table \ref{tab:fiber_comparison}). When these fibers are directed into a slit arrangement, the total height does not surpass $384\mu m$.

\subsection{Hexagonal Fiber}

\begin{figure}[htbp]
  \centering
  \begin{subfigure}[b]{0.30\linewidth}
    \includegraphics[width=\linewidth]{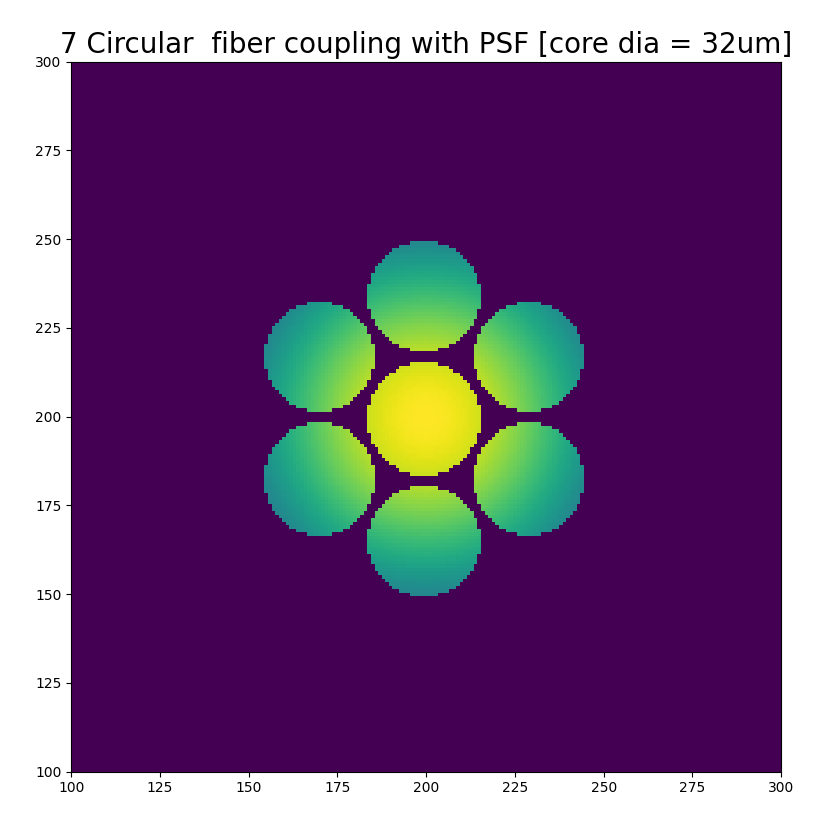}
    \caption{Circular Fiber Arrangement }
    \label{fig:sub1}
  \end{subfigure}
  \begin{subfigure}[b]{0.32\linewidth}
    \includegraphics[width=\linewidth]{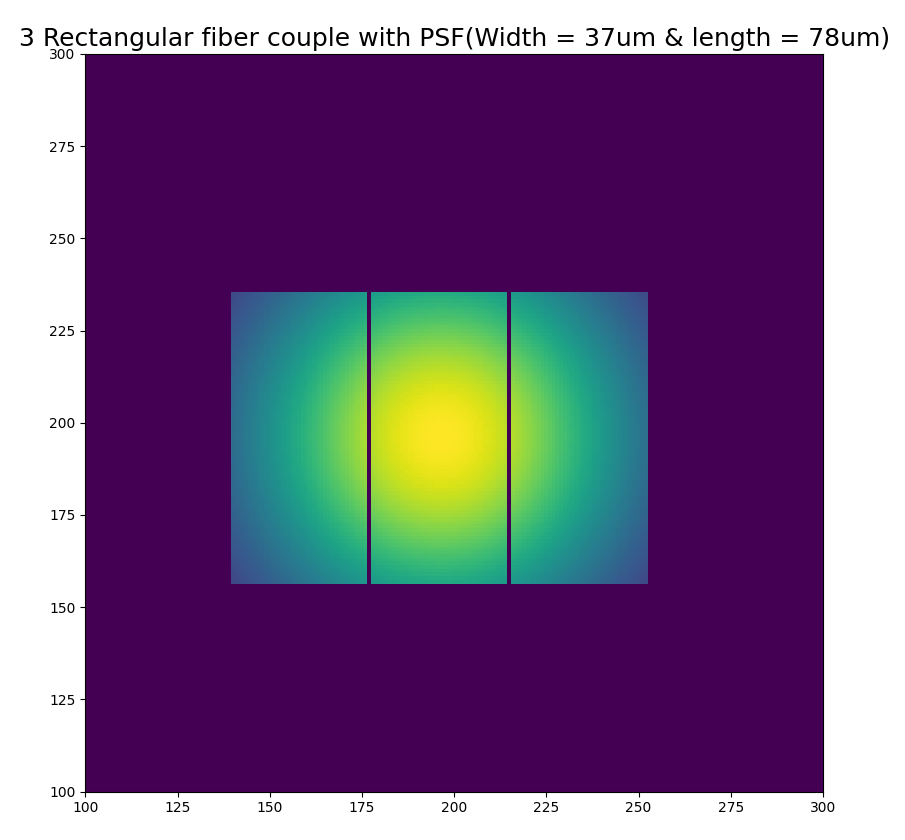}
    \caption{Rectangular Fiber Arrangement}
    \label{fig:sub2}
  \end{subfigure}
  \begin{subfigure}[b]{0.30\linewidth}
    \includegraphics[width=\linewidth]{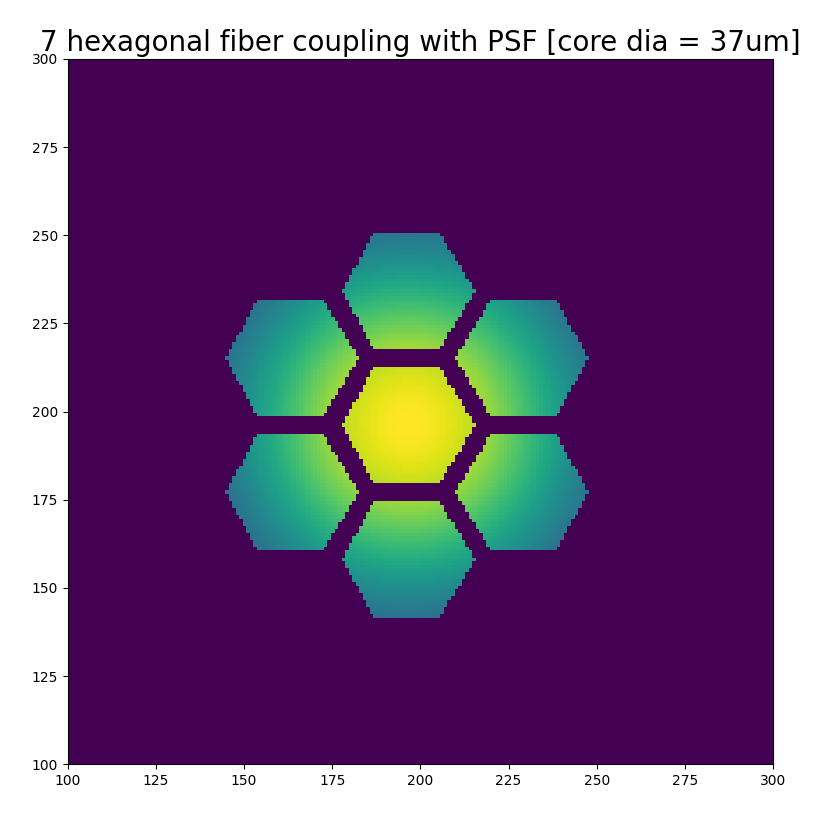}
    \caption{Hexagonal Fiber Arrangement}
    \label{fig:sub3}
  \end{subfigure}
  \caption{Different types Fiber Arrangement on Prime Focus of VBT for Increasing the Coupling Efficiency}
  \label{fig:fiber arrangment}
\end{figure}

We also explored the possibility of using 7 hexagonal fibers arranged in a focal plane Fig.~\ref{fig:sub3}. Once again, the output from these fibers will form a slit for the spectrograph. Accounting for the necessary magnification, we calculate the diameter of each fiber using Equation.~\ref{eq:height_fiber}:

\vspace{-3mm}
\begin{equation}
    \text{Fiber diameter (core + cladding)} = \frac{384}{7 \times 1.6 \times 0.86} = 39 \mu m
    \label{eq:7hexa}
\end{equation}

From Equation \ref{eq:7hexa}, we find that the maximum fiber diameter we can use in our simulation is 39 $\mu m$. Therefore, we will use a 37 $\mu m$ core diameter with a 2 $\mu m$ cladding diameter fiber.

\vspace{-2mm}

 \section{Throughput Calculation on Detector using New Fiber Arrangement}

\begin{figure}[htbp]
  \centering
  \begin{subfigure}[h]{0.47\linewidth}
    \centering
    \includegraphics[width=\linewidth]{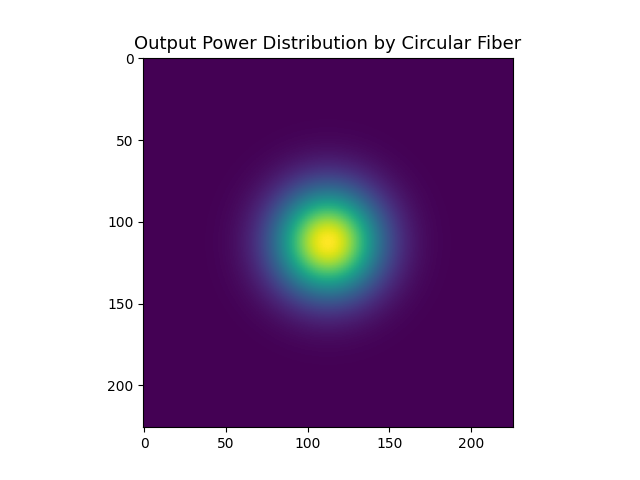}
    \caption{Output Power Distribution by Circular Fiber}
    \label{fig:circul_psf}
  \end{subfigure}
  \hspace{5mm} 
  \begin{subfigure}[h]{0.47\linewidth}
    \centering
    \includegraphics[width=\linewidth]{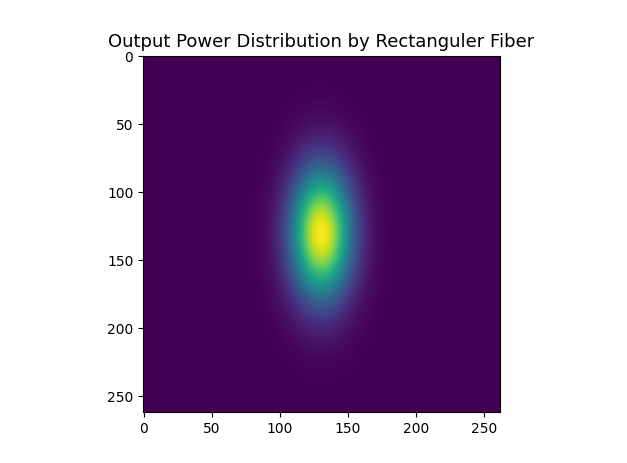}
    \caption{Output Power Distribution by Rectangular Fiber}
    \label{fig:rect_psf}
  \end{subfigure}
  \caption{Power distribution profiles for different fiber shapes}
  \label{fig:psf_shapes}
\end{figure}

After maintaining sufficient inter-order separation, the diameter of different core-shaped fibers falls within the range of 50 to 60 $\mu m$ after magnification (see Table \ref{tab:fiber_comparison}). All fibers are aligned in a slit fashion. We did a simulation using three different types of fibers (see Table \ref{tab:fiber_comparison}). The circular and hexagonal fibers exhibit output power distribution functions with Gaussian symmetry, resulting in a symmetrical Gaussian function as illustrated in Fig.~\ref{fig:circul_psf}. However, in the case of the rectangular fiber, the output power distribution function is elongated in the height direction, corresponding to the slit height as illustrated in Fig.~\ref{fig:rect_psf}.

\vspace{-3mm}
\begin{table}[h]
    \centering
    \caption{\textbf{Fiber Arrangement}}
    \begin{tabular}{|l|c|c|}
        \hline
        \textbf{Fiber Type} & \textbf{Core Diameter ($\mu$m)} & \textbf{Diameter after Magnification ($\mu$m)}  \\
        \hline
        Circular Fiber & 32  & 32 $\times$ 1.6 = 51  \\
        Rectangular Fiber & 37 $\times $78& 37 $\times$ 1.6 = 59  \\
        Hexagonal Fiber & 37  & 37 $\times$ 1.6 = 59  \\
        \hline
    \end{tabular}
    \label{tab:fiber_comparison}
\end{table}

We observe that the after-magnification diameter of the fiber is less than 60 $\mu m$, which is the slit width required for $R \approx 72,000$. For achieving higher $R \approx 100,000$, a slit width of 30 $\mu m$ is required. Since the minimum resolution achievable using this fiber bundle setup is $R \approx 72,000$, we will employ the spectral binning method to achieve $R \approx 27,000$. To assess the throughput efficiency at $R \approx 72,000$ and $100,000$, we summarized the results obtained from the simulation of the new fiber arrangements in Tables \ref{tab:throu_72000} and \ref{tab:throu_100000}, respectively.

\begin{table}[h]
    \centering
    \caption{\textbf{Throughput Calculations For  R$\approx$ 72,000}}
    \begin{tabular}{|p{2.8cm}|p{2.3cm}|p{2.3cm}|p{2cm}|p{2.3cm}|p{2.3cm}|}
        \hline
        \textbf{Fiber Type}    & \textbf{Fiber Coupling on Prime Focus} &  \textbf{Fiber Transmission}  &   \textbf{Slit Coupling} & \textbf{Total throughput by Optics} & \textbf{Total Throughput to detector}  \\
         \hline
         
        Circular Fiber & 0.39 & $0.98^7$  & 0.99 & 0.58 & 0.19    \\
         \hline
    
        Rectangular Fiber  & 0.57 & $0.98^3$  & 0.99 & 0.58 & 0.30    \\
        \hline
        Hexagonal Fiber  & 0.44  & $0.98^7$  & 0.99 & 0.58 & 0.22    \\
        \hline
    \end{tabular}
    \label{tab:throu_72000}
\end{table}

\begin{table}[h]
    \centering
    \caption{\textbf{Throughput Calculations For R$\approx$ 100,000 (slit width = $30\mu m$ )}}
    \begin{tabular}{|p{2.8cm}|p{2.3cm}|p{2.3cm}|p{2cm}|p{2.3cm}|p{2.3cm}|}
        \hline
        \textbf{Fiber Type}    & \textbf{Fiber Coupling on Prime Focus} &  \textbf{Fiber Transmission}  &   \textbf{Slit Coupling} & \textbf{Total throughput by Optics} & \textbf{Total Throughput to detector}  \\
         \hline
         
        Circular Fiber & 0.39 & $0.98^7$  & 0.51 & 0.58 & 0.10    \\
         \hline
    
        Rectangular Fiber  & 0.57 & $0.98^3$  & 0.67 & 0.58 & 0.20    \\
        \hline
        Hexagonal Fiber  & 0.44  & $0.98^7$  & 0.45 & 0.58 & 0.10    \\
        \hline
    \end{tabular}
    \label{tab:throu_100000}
\end{table}

The total throughput from optics is computed using Equation \ref{eq:optics_thro} and fiber transmission of 98 \% for all fibers. In both Tables \ref{tab:throu_72000} and \ref{tab:throu_100000}, the fiber coupling indicates the amount of light coupled by the fiber when densely packed in prime focus as illustrated in Fig.~\ref{fig:fiber arrangment}. It's evident that the rectangular fiber couples more light compared to the circular and hexagonal arrangements. Although the current HRES fiber-fed setup exhibits a higher fiber coupling efficiency (capturing 55 \% of the light see Fig.~\ref{fig:fiber_coupl}), compared to the new circular and hexagonal fiber arrangements (see Table \ref{tab:table1}, \ref{tab:throu_72000}, and \ref{tab:throu_100000}) in prime focus. The total throughput at the HRES detector is higher with the new fiber arrangements due to the improved slit coupling efficiency achieved by the new fiber arrangement, which is nearly double in the case of circular and hexagonal fibers, and triple in the case of rectangular fibers for R $\approx$ 72,000 and 100,000.
 
\section{Conclusion}

The maximum light collected from the telescope, feeding to the high-resolution spectrographs through fibers, presents a challenge. Our study addresses this through the theoretical analysis of different fiber arrangements to enhance the coupling efficiency for the VBT HRES.  We studied three (circular, hexagonal, rectangular) fiber arrangements to mitigate flux loss and improve coupling efficiency, retaining the spectral resolutions of HRES and maintaining sufficient inter-order separation. The circular and hexagonal fiber arrangements demonstrate that the flux throughput efficiency is nearly double that of the current HRES single fiber setup. Interestingly, with the rectangular fiber configuration, the throughput can be increased thrice for R~ 72,000 and 100,000 (see Table~\ref{tab:table1},~\ref{tab:throu_72000}, and \ref{tab:throu_100000}). We note that in all these new fiber arrangements, the spectral resolution of $R \approx 27,000$ can be achieved through the spectral binning method. The throughput efficiency from rectangular fiber arrangement is comparable to the present setup of HRES. However, the other arrangements show a 10 \% less throughput (see Table~\ref{tab:table1}, and ~\ref{tab:throu_72000}). Overall, our study suggested that the rectangular fiber setup offers the best performance to enhance the coupling efficiency for VBT HRES.

\bibliography{report} 
\bibliographystyle{spiebib} 

\end{document}